\documentclass[12pt]{iopart}
\usepackage{amssymb}
\usepackage{amstext}
\usepackage{enumerate}
\renewcommand\epsilon{\varepsilon}
\newtheorem{theo}{Theorem}[section]

\newtheorem{prop}{Proposition}[section]
\newtheorem{lemma}{Lemma}[section]

\newcommand{\id}{{1 \mskip -5mu {\rm I}}}

\begin{document}

\title[Recursive Pair Substitutions]{Non Sequential Recursive Pair Substitution: Some
Rigorous Results}

\author{Dario Benedetto$^1$, Emanuele Caglioti$^1$ and Davide Gabrielli$^2$}
\address{$^1$  Dipartimento di Matematica, Universit\`a di Roma 
``La Sapienza'',
P.le A. Moro 2, 00185 Roma, Italy. e-mail:
benedetto@mat.uniroma1.it ; caglioti@mat.uniroma1.it}
\address{$^2$ Dipartimento di Matematica, 
Universit\`a dell'Aquila, via Vetoio Loc. Coppito, 67100 L'Aquila, Italy. e-mail: gabriell@univaq.it}

\begin{abstract}
We present rigorous results on some open questions on NSRPS,
non sequential
recursive pairs substitution method
(see Grassberger in \cite{G}). In particular, starting from the action
of NSRPS on finite strings we define a corresponding natural action
on measures and we prove that the iterated
measure becomes asymptotically Markov. This certify the
effectiveness of NSRPS as a tool for data compression and entropy
estimation.
\end{abstract}



\section{Introduction}
%definizione con esempi delle NSRPS
We consider here a suitable non sequential recursive pair
substitution method (NSRPS) which has been proposed by
Jimenez-Monta\~no, Ebeling and others \cite{t}. This method has been
studied and precisely defined by P. Grassberger as a tool for data
compression and entropy estimation \cite{G}. He deduced some
important properties of the method and used it to estimate the
entropy of the written English.

In particular the results found in \cite{G} and the conjectures made
therein are the main motivation for this paper.

%pezzo di emanuele**************************************************
%***************************************************************

Data compression is one of the most interesting research fields in
Information Theory both from the applied and from the theoretical
viewpoint. In particular data compression algorithms provide a
powerful tool for the measure of the entropy and more in general for
the estimation of complexity of a sequence. The first algorithms
(Shannon-Fano, Huffman, see for example \cite{A}, \cite{S}) 
were based on the suitable coding of single
characters, or of strings of a fixed and small number of characters.
A great improvement in the field of data compression has been given
by the dictionary-based compression methods LZ77 \cite{ZL1}, LZ78
\cite{ZL2} and LZW \cite{W} in which variable-length strings are suitably encoded. In
particular in LZ78 a sequence is encoded as a list of phrases.
Initially the phrases coincide with the characters and then any new
phrase is obtained sequentially by adding a character to one of the
existing phrases. The NSRPS method we are going to study here, even
if different in many respects from this dictionary methods, has some
similarity with LZ78 and in particular  with a variation of LZ78
which has been recently proposed \cite{CaSToRe}.

%fine pezzo emanuele***********************
%**********************************************

The NSRPS method works in the following way. Let us consider a
sequence $\underline s^0$ built with the characters of a finite
alphabet $A=\{a_0,...,a_{m-1}\}.$ For any given $i,j$ let $n_{ij}$
be the number of non-overlapping occurrences of the string $a_i a_j$
in $\underline s^0,$ and let be $i_0,j_0$ the pair (or one of the
pairs) for which $n_{ij}$ is maximum. Now let us define a new
sequence $\underline s^1$ obtained from $\underline s^0$ by
substituting any occurrence of the pair $a_{i_0}a_{j_0}$ with a new
symbol $a_m.$ The new sequence is shorter than the previous one and
its alphabet has one character more. Then starting from $\underline
s^1$ we define a new sequence $\underline s^2$ with the same
procedure, et cetera. We call a single step of NSRPS a ``pair
substitution'' (the one for example that transforms
$\underline{s}^0$ into $\underline{s}^1$).

For sake of clearness let us consider two specific examples when the
initial sequence is binary.
For first let us consider the case in which
$$\underline s^0=0010101010001001010101110101.....$$
and we substitute $01$ with the new character $2.$
We obtain
$$\underline s^1=02222002022221122.....$$
As said above the sequence $\underline s^1$ is shorter then
$\underline s^0$. In
particular, denoting with $|\underline s|$
the length of a generic sequence $\underline s,$
we have
$$|\underline s^1|=|\underline s^0|-\#\{01\subseteq \underline
s^0\},$$ where $\#\{01\subseteq \underline s^0\}$ is the number of
times we find $01$ in the string $\underline s^0$. Dividing by
$|\underline s^0|$
$$\frac{|\underline s^1|}{|\underline s^0|}=1-
\frac{\#\{01\subseteq \underline s^0\}}{|\underline s^0|}.$$
We always work
with sequences extracted by an ergodic measure $\mu$. Then
taking the limit as
$|\underline s^0|\to\infty$
we get, for almost all
sequences $\underline s^0,$ that
\begin{equation}
\frac1Z :=\lim_{|\underline s^0|\to\infty} \frac{|\underline
s^1|}{|\underline s^0|}=1-\mu(01). \label{mancava}
\end{equation}

Another important fact to notice is that the transformation is
invertible (see Section \ref{sez:stringhe}), and then the amount of
information of the two sequences is the same (see Section
\ref{sez:entropie}). Therefore, if $h(\underline s)$ is the entropy
per character of $\underline s$:
$$h(\underline s^0)=\frac{h(\underline s^1)}{Z}.$$

The second example we consider is when the pair to be substituted
is made by two equal characters.
Let us consider the sequence
$$1001100100000011001000001000010001\dots $$
and let us substitute $00$ with $2.$
We found the new sequence
$$12112122211212201221201\dots$$
The main difference with the case considered before is the fact that
in this case we do not substitute with $2$ all the pairs of
consecutive $0$ in $\underline s^0.$ For instance $1001\to 121,$ but
$10001\to 1201$. It is easy to deduce that in this case
(\ref{mancava}) changes in
\begin{equation}
\label{intro2}
\frac1Z=1-\mu(00)+\mu(000)-\mu(0000)+\mu(00000)-......
\end{equation}
This example shows that under a NSRPS the probabilities of
strings can behave in a complicate way.
In spite of this fact, the
substitution process transform a Markov sequence in a Markov sequence,
as proved by Grassberger in \cite{G}.

In general, if the starting sequence is not Markov it does not
becomes Markov after a finite number of transformations.
Nevertheless it is reasonable to expect that the sequences tends to
become Markov as the number of transformations tends to infinity.
This is exactly what was conjectured in \cite{G} and what we prove
here.

More precisely the main facts we prove are the following.

In any pair substitution the conditional entropy $h_1$ (i.e. the
entropy of a character conditioned to the previous character),
suitably normalized, does not increase. If the process is already
Markov then it stays constant (truly, there are other rare cases in
which $h_1$ stays constant, see Section \ref{sez:esempio} and
Section \ref{appendice:entropie}).

This is a
general property of the pair transformations and holds true whatever
is the substitution made. An immediate corollary of this fact is that
Markov sequences are transformed in Markov sequences.

As the number of transformations goes to $\infty$ and also the
inverse of the average shortening $Z$ diverges, the (suitably
normalized) conditional entropy $h_1$ tends to the entropy  of the
sequence. In this sense
 we prove that in the limit the process becomes Markov.
In particular this is the case if any time we substitute the pair of
characters which maximizes the number of nonoverlapping occurrences.
This condition is not strictly necessary but, as we shall see in
Section \ref{sez:esempio}, not for all the sequences
of substitutions it holds the result.

The paper is organized as follows.

In section $2$ we will fix notations and give some preliminary
results. In particular we will discuss how pair substitutions act on
strings and give a natural definition of a corresponding action on
ergodic measures.

In section $3$ we will state results on how pair substitutions act
on entropies.

In section $4$ we prove the main result of the paper.

In section $5$ we discuss some examples.

In section $6$ we give some concluding remarks.

In sections $7,\,8$ we collect technical results on measures
and entropies transformations under the action of a pair substitution,
respectively.

\section{How  pair substitutions act on strings and measures}
\label{sez:stringhe}

\subsection{Strings}
Given an alphabet $A$ we denote with $A^*:=\cup_{k=1}^{+\infty}A^k$
the set of finite words in the alphabet $A$. Elements of $A^*$ are
indicated with underlined lower case Latin letters
$\underline{w},\underline{x},etc.$ The same notation will be used
also for infinite (elements of $A^{\mathbb N}$) and double infinite
words (elements of $A^{\mathbb Z}$). An element $\underline{w}$ has
length $|\underline{w}|$ and, if $|\underline w|=k$, it is also
indicated with $w_1^{k}:= w_1\dots w_{k}:=(w_1,\dots,w_k)$.

\vskip8pt Let us consider $x,y\in A$ (including $x=y$),
$\alpha\notin A,$ and $A'=A\cup\{\alpha\}.$ A {\it  pair
substitution} is a map $G=G_{xy}^{\alpha}:A^*\to {A'}^*$ which
substitutes ordinately the occurrence of $xy$ with $\alpha$. More
precisely $G\underline{w}$ is defined by substituting in
$\underline{w}$ the first occurrence from the left of $xy$ with
$\alpha,$ and then repeating this procedure till the end of string.

We define also the map $S=S_{\alpha}^{xy}:A'^*\to A^*$, which
 acts on the words
$\underline{z}\in A'^*$ substituting any occurrence of the symbol $\alpha$
with the pair $xy$.
\vskip5pt
Notice that the map $G$ is injective and not surjective while the map
$S$ is surjective and not injective. Notice also that
$S|_{G(A^*)}=G^{-1}$, i.e.
\begin{equation}
\label{mapSG}
S(G(\underline w)) = \underline w\ \text{ for any }\ \underline
w \in A^*.
\end{equation}

We remark that these definitions work also in the case of infinite
sequences $\underline{w}\in A^{\mathbb N}$ and $\underline{z}\in
A'^{\mathbb N}$.

It is easy to see that the set of admissible words $G(A^*)$
is a subset of $A'^*$ which can be described by constraints on
consecutive symbols: in the case  $xy\to \alpha$, with $x\neq y$,
$G(A^*)$ consists of the strings of $A'^*$ in which does not
appear the pair $xy$; in the case $xx\to \alpha$,
$G(A^*)$ consists of the strings of $A'^*$ in which do not
appear the pairs $xx$ and $x\alpha$.
An important fact is that after
the application of more pair substitutions,
the set of admissible words remains
described by constraints on
consecutive symbols. This follows from the fact that a pair
substitution
maps pair constraints in pair constraints, as stated in the
following theorem.

\begin{theo}
\label{teo:vincoli}
Let $\left\{ V_{a,b} \right\}_{a,b\in A}$ be a matrix with 0--1
valued elements
(the constraint matrix), and
let $A^*_V$ be the subset of $A^*$ whose elements $\underline w$
verify
$$
\prod_{i=1}^{|\underline w|-1}V_{w_i,w_{i+1}}=1,
$$
( $A^*_V$ is the set of admissible strings with respect to
the pair constraints given by $V$).

\noindent
There exists a constraint matrix $V'$ with index in $A'$
such that
$$G(A^*_V) = A'^*_{V'}.$$
\end{theo}
The proof follows from direct inspection. Here we only
write $V'$ in
terms of $V$.
Let $z,w\in A\setminus \{x,y\}$:
the values of the elements of $V'$ are given by the following tables:
\vskip10pt

\begin{tabular}{l|r|r|r|r|}
if $x\neq y$
         & $x$       & $y$       & $w$       & $\alpha$   \\
\hline
$x$      & $V_{x,x}$ & 0         & $V_{x,w}$ & $V_{x,x}$  \\

\hline
$y$      & $V_{y,x}$ & $V_{y,y}$ & $V_{y,w}$ & $V_{y,x}$  \\
\hline
$z$      & $V_{z,x}$ & $V_{z,y}$ & $V_{z,w}$ & $V_{z,x}$  \\
\hline
$\alpha$ & $V_{y,x}$ & $V_{y,y}$ & $V_{y,w}$ & $V_{y,x}$  \\
\end{tabular}\hskip60pt
\begin{tabular}{l|r|r|r|}
if $x=y$
         & $x$       & $w$       & $\alpha$   \\
\hline
$x$      & 0         & $V_{x,w}$ & 0          \\
\hline
$z$      & $V_{z,x}$ & $V_{z,w}$ & $V_{z,x}$  \\
\hline
$\alpha$ & 1         & $V_{x,w}$ & 1          \\
\end{tabular}

\vskip5pt
\noindent
Note that these expressions hold if
$V_{x,y} = 1$ and $V_{x,x}=1$ respectively; otherwise, and this is a non
interesting case, $G(A^*_V)=A^*_V$.

\subsection{Measures}
 We indicate
with $\mathcal E(A)$ the set of ergodic stationary measures
on $A^{\mathbb Z}$, the only measures we are interested in.
If $\mu \in \mathcal E(A)$ we use the shorthand notation
$\mu(\underline w)$ to indicate the value of the $|\underline w|$-marginals
of $\mu$ on the sequence $\underline w$.

The maps $G_{xy}^{\alpha}$ and $S_{\alpha}^{xy}$ induce
the maps $\mathcal G=\mathcal G_{xy}^{\alpha}:\mathcal E(A) \to
\mathcal E(A')$ and
 $\mathcal S=\mathcal S_{\alpha}^{xy}:\mathcal E(A') \to
\mathcal E(A)$ in the following natural sense.
Let $\mu\in \mathcal E(A)$ and
 $\underline w\in A^{\mathbb N}$ be a frequency typical sequence with respect to
$\mu$,  and let $\nu\in \mathcal E(A')$ and $\underline z\in A'^{\mathbb N}$
 be a frequency typical sequence with respect to
$\nu$. The sequence $G\underline{w}$ is typical for an ergodic measure
that we call $\mathcal G\mu$ and the sequence $S\underline z$
is typical for an ergodic measure that we call $\mathcal S\nu$.

More precisely,
denoting the number of
occurrences of a subword
$\underline{s}$ in $\underline{r}$ with
$
\sharp\left\{\underline{s}\subseteq
\underline{r}\right\}:=\sum_{i=1}^{|\underline{r}|-|\underline{s}|+1}\id
(r_i^{i+|\underline{s}|-1}=\underline{s})
$,
where $\id$ is the characteristic function,
it holds
\begin{theo}
\label{teo:mis}
Let $\underline s\in A'^*$ then
\begin{equation}
\label{def:misG} \mathcal G\mu(\underline s) := \lim_{n\to +\infty}
\frac {\sharp \{\underline s \subseteq G(w_1^n) \}}
{|G(w_1^n)|}\
\end{equation}
exists and is constant $\mu$ almost everywhere in $\underline w$, moreover
$\left\{\mathcal G\mu(\underline s)\right\}_{\underline{s}\in A'^*}$
are the marginals of an ergodic measure on $A'^{\mathbb{Z}}$.
\vskip5pt
\noindent
In analogous way, let $\underline r\in A^*$;
then
\begin{equation}
\label{def:misS}
\mathcal S\nu(\underline r) := \lim_{n\to +\infty}
\frac {\sharp \{\underline r \subseteq S(z_1^n) \}}
{|S(z_1^n)|}\
\end{equation}
exists and is constant $\nu$ almost everywhere in $\underline z$, moreover
$\left\{\mathcal{S}\nu(\underline r)\right\}_{\underline{r}\in A^*}$
are the marginals of an ergodic measure on $A^{\mathbb{Z}}$.
It holds
\begin{equation}
\label{misSG}
\mathcal S_{\alpha}^{xy}\mathcal G_{xy}^{\alpha} \mu = \mu.
\end{equation}
\end{theo}
In Section \ref{appendice:mis} we give the proof of the theorem and
of the following propositions (which we use
for the main theorem in Section \ref{sez:main});
moreover from
(\ref{def:misG}) and (\ref{def:misS}) we write
the explicit expressions of $\mathcal G\mu$ and $\mathcal
S\nu$ in terms of $\mu$ and $\nu$ respectively.

\vskip5pt
\begin{prop}
\label{teo:ZW}
Let $Z_{xy}^{\mu}$ be the inverse of the
 mean shortening, with respect to
$\mu$, of a string under the action of
$G_{xy}^{\alpha}$
and let $W=W_{\alpha}^{\nu}$ be the mean lengthening, with respect to
$\nu$, of a string under
the action of $S_{\alpha}^{xy}$.

\begin{equation}
\label{def:Z}
\text{If }x\neq y\ \ Z_{xy}^{\mu}:=
\lim_{n\to +\infty}  \frac n{|G(w_1^n)|} =
\frac 1{1 - \mu(xy)}\ \ \ \ (\mu \text{ a.e. in }\underline w).
\end{equation}
\begin{equation}
\label{def:Zxx}
{Z_{xx}^{\mu}} := \lim_{n\to +\infty}  \frac n{|G(w_1^n)|} =
\frac 1{
1 - \sum_{k=2}^{+\infty} (-1)^k \mu(\underline x^k)}
\ \ \ \ (\mu \text{ a.e. in }\underline w),
\end{equation}
where $\underline x^k$ is the sequence of $k$ times $x$.
\begin{equation}
\label{def:W}
W_{\alpha}^{\nu} :=
\lim_{n\to +\infty}  \frac {|S(z_1^n)|}{n} = 1 + \nu(\alpha)
\ \ \ \ (\nu \text{ a.e. in }\underline z).
\end{equation}
Moreover
\begin{equation}
\label{Z=W}
W_{\alpha}^{\mathcal G_{xy}^{\alpha}\mu} = Z_{xy}^{\mu}
\end{equation}
\end{prop}
\noindent

\begin{prop}
\label{prop:S}
Let $\underline r\in A^*$, the value of
$\mathcal S\nu(\underline r)$ depends only on
the values of $\nu(\underline s)$ with
$|\underline s|\leq |\underline r|$
\end{prop}
We remark that this assertion if false for $\mathcal G\mu$,
and, in the case of $x=y$,
$\mathcal G\mu(\underline s)$ can involve the probability
of infinitely many strings of increasing lengths (see Eq.s (\ref{inv2})).

\begin{prop}
\label{propo:invS}(Invertibility of $\mathcal S \nu$)

\noindent
If $\nu\in\mathcal E(A')$ respects the pair constraints given by $G$,
i.e. for $\underline z\in A'^*$
$$\nu(\underline z) = 0\ \ \text{ if }\ \underline z \notin G(A^*),$$
then
$$\nu = \mathcal G \mathcal S \nu.
$$
\end{prop}

\section{How  pair substitutions act on the entropy per symbol}
\label{sez:entropie}
Given $\mu\in \mathcal E(A)$, $n\geq 1$, and indicating
with $\log$ the base 2 logarithm function,
$$
\begin{array}{l l}
H_n(\mu):=-\sum_{ |\underline z|=n } \mu(\underline z)
\log \mu(\underline z) & \text{ is the  n--block entropy,}\\
h_n(\mu):=H_{n+1}(\mu)-H_n(\mu) &
\text{ is the n--conditional entropy,}\\
h(\mu) := \lim_{n\to +\infty} \frac {H_n(\mu)}{n} =
\lim_{n\to +\infty} h_n(\mu) &\text{ is the entropy of }\mu.
\end{array}
$$
It holds:
\begin{equation}
\label{verso_entropie}
h(\mu) \le \dots \le h_{j}(\mu)
 \le h_{j-1}(\mu) \le \dots \le h_1(\mu) \le H_1(\mu).
\end{equation}
Denoting with $\mu(\underline z|\underline w):=
{\mu(\underline w\,\underline z)}/{\mu(\underline w)}$
the conditional probabilities, we say that
$\mu$ is a $k-$Markov measure if for any $n>k$,
$\underline w\in A^n$ and $a\in A$,
$\mu(a|w_1^n) = \mu(a|w_{n-k+1}^n)$.
In this case $h(\mu)=h_j(\mu)\ \ \ \forall j\geq k$.
We remark that
%$h_k(\mu)$ is a non increasing sequence and that
$h(\mu)=h_k(\mu)$ implies that $\mu$ is a $k-$Markov measure.

We collect here some results on how entropies transform
under the action of $\mathcal{G}$. Proofs are postponed to the technical
Section \ref{appendice:entropie}.

We will use the shorthand $Z=Z_{xy}^{\mu}$, and sometimes
$Z^{\mu}=Z_{xy}^{\mu}$ when
%the pair $xy$ is not important and
we need to stress the reference measure.

\begin{theo}
\label{teo:h}
\noindent
\begin{equation}
\nonumber
h(\mathcal G\mu) =  Z {h(\mu)}.
\end{equation}
\end{theo}
In fact the information amount of the string $\underline w$
is the same of the string $G(\underline w)$.
%The proof of
%this theorem is postponed to section \ref{appendice:entropie}.
%(a more formal proof can be obtained with little modifications of
%the classical Shannon-McMillan theorem).

\begin{theo}
\noindent
\begin{equation}
\label{th:teoh1}
h_1(\mathcal G\mu) \leq Z h_1(\mu)
\end{equation}
\label{teo:h1}
Moreover, if $\mu$ is a $1-$Markov measure
$\mathcal G \mu$ is a $1-$Markov measure.
\end{theo}

\noindent
Let us notice here that
the second assertion is a consequence of the
first: if $\mu$ is a $1-$Markov measure
\begin{equation}
h(\mathcal G\mu) \leq  h_1(\mathcal G\mu) \leq Z h_1(\mu) =
Zh(\mu) = h(\mathcal G\mu).
\nonumber
\end{equation}
Then $h_1(\mathcal G\mu)= h(\mathcal G\mu)$; this implies that
 $\mathcal G \mu$ is a $1-$Markov measure.

This theorem can also be generalized.

\begin{theo}
\noindent
\label{teo:hk}
\begin{equation}
\nonumber
h_k(\mathcal G \mu) \leq Z h_k(\mu),
\end{equation}
and $\mathcal G$ maps $k-$Markov measures
in $k-$Markov measures.
\end{theo}

\section{The main result}
\label{sez:main}
Theorem \ref{teo:h1} asserts,
roughly speaking, that the amount of information
of $G(\underline w)$, which is equal to that of $\underline w$,
is more concentrated on the pairs of symbols, with respect to the
case of the original string $\underline w$.
This fact suggests that a sequence of
pair substitutions can transfer all the information in
the distributions of the pairs of symbols.
To formalize this assertion, let us define
recursively:

the alphabets $A_{_N}= A_{_{N-1}} \cup \{\alpha_{_N}\}$ where
$\alpha_{_N} \notin A_{_{N-1}}$, with $A_0 = A$;

the maps $G_{_N}=G_{x_{_N}y_{_N}}^{\alpha_{_N}}: A_{_{N-1}}^*
\to A_{_N}^{*}$,
where
$x_{_N},\,y_{_N} \in A_{_{N-1}}$;

the corresponding maps
$\mathcal G_{_N}=\mathcal G_{x_{_N}y_{_N}}^{\alpha_{_N}}$,
$S_{_N}=S_{\alpha_{_N}}^{x_{_N} y_{_N}}$,
$\mathcal S_{_N}=\mathcal S_{\alpha_{_N}}^{x_{_N}y_{_N}}$;

the measures
$\mu_{_N}=\mathcal G_{_N}\mu_{_{N-1}}$, with $\mu_0 = \mu$;

the normalization $Z_{_N}= Z_{x_{_N}y_{_N}}^{\mu_{_{N-1}}}$;

the composed maps
$$\begin{array}{l l}
\overline{G}_{_N}=G_{_N}\circ\cdots\circ G_1,\
&\overline{\mathcal G}_{_N}=\mathcal G_{_N}\circ \cdots \circ \mathcal G_1,\\
\overline{S}_{_N}=S_1\circ\cdots\circ S_{_N},\
&\overline{\mathcal S}_{_N}=\mathcal S_1\circ \cdots\circ \mathcal S_{_N};\\
\end{array}$$

the corresponding normalization
$\overline{Z}_{_N}= Z_{_N}Z_{_{N-1}}
\cdots Z_1$
(when we need to specify
the initial measure we will use the symbol $\overline{Z}_{_N}^{\mu}$).

\vskip5pt
\noindent
In \cite{G} the author chose at any step the pair of symbols
with the
 maximum of the frequency of non-overlapping occurrences.
This fact assures the divergence of $\overline{Z}_{_N}$ as we will
prove using Theorem \ref{teo:h1}.
\begin{theo}
\label{teo:divZ} If at any step $N$ the pair $x_{_N}y_{_N}$ is the
pair of maximum of  frequency of non-overlapping occurrences between
the pairs of symbols of $A_{_{N-1}}$ then
\begin{equation}
\nonumber
\lim_{N\to +\infty} \overline{Z}_{_N} = +\infty
\end{equation}
\end{theo}
In this case the hypothesis of the following (main)
theorem is satisfied.
\begin{theo}
\label{teo:main}
If
\begin{equation}
\label{divZ}
\lim_{N\to +\infty} \overline{Z}_{_N} = +\infty
\end{equation}
then
\begin{equation}
\label{tesi}
h(\mu) = \lim_{N\to +\infty} \frac {h_1(\mu_{_N})}{\overline{Z}_{_N}}
\end{equation}
\end{theo}

\noindent
{\bf Proof of Th. \ref{teo:divZ}}

Let $p_{_N}$ the maximum of probability $\mu_{_{N-1}}$
on the pair of symbols of $A_{_{N-1}}$. From the definition
of $Z_{_N}$ it follows that
$$\overline{Z}_{_N} \geq \overline{Z}_{_{N-1}}
\left(1 + \frac {p_{_N}}2\right),$$
(the factor 2 appears for the case of substitution of two equal symbols).
We can estimate $p_{_N}$ with
$$p_{_N} \geq 2^{-H_2(\mu_{_{N-1}})},$$
where $H_2(\mu_{_{N-1}}) = - \sum_{a,b \in A_{_{N-1}}}
\mu_{_{N-1}}(ab) \log \mu_{_{N-1}}(ab)$ is the 2-block entropy.
Using Th. \ref{teo:h1} and that $H_1(\mu_{_{N-1}})\leq \log (N-1+|A|)$,
with $|A|$ the cardinality of $A$:
$$H_2(\mu_{_{N-1}})=h_1(\mu_{_{N-1}}) + H_1(\mu_{_{N-1}})
\leq \overline{Z}_{_{N-1}} h_1(\mu) + \log(N-1+|A|).$$
Then
$$\frac {\overline{Z}_{_N}}{\overline{Z}_{_{N-1}}} \geq 1
+ \frac {2^{-\overline{Z}_{_{N-1}} h_1(\mu)}}{2(N-1+|A|)}.$$
The sequence $\overline{Z}_{_N}$ is increasing; by absurd,
if $\overline{Z}_{_N}$ tends
to a constant, from the previous equation
$\overline{Z}_{_N}/\overline{Z}_{_{N-1}} \geq 1 + c/(N-1)$,
but this implies $\overline{Z}_{_N}\to +\infty$.

\vskip5pt \noindent {\bf Remark:} this proof is also valid in the
more general case we choose $x_{_N}y_{_N}$ in such a way that
$$\mu_{_{N-1}}(x_{_N}y_{_N})\geq c p_{_N},$$
where $c$ is a constant independent on $N$.

\vskip10pt
\noindent
{\bf Proof of Th. \ref{teo:main}}

For the composition $\overline{ S}_{_N}$ it holds
$$
\overline{S}_{_N}(s_1^n)=\overline{ S}_{_N}(s_1)\dots \overline{ S}_{_N}(s_n),
$$
where $\overline{S}_{_N}(s_i)$ are words in the original alphabet
$A$. Consider $\underline r\in A^*$, $|\underline r|=k$ and
$\underline{s}$ a typical string for $\mu_{_N}$.
\begin{eqnarray}
\mu(\underline{r})=  \lim_{n\to \infty}
\frac{\sharp\left\{\underline{r}\subseteq \overline{S}_{_N}(s_1^n)
\right\}}{|\overline{S}_{_N}(s_1^n)|}
\nonumber
 = \lim_{n \to \infty}
\frac{\sharp\left\{\underline{r}
\subseteq \overline{S}_{_N}(s_1)\dots
\overline{S}_{_N}(s_n)\right\}}{|\overline{S}_{_N}(s_1^n)|}
\nonumber
\end{eqnarray}
Notice that
\begin{equation}
\hskip-30pt
\begin{array}{ll}
&\sharp\left\{\underline{r}
\subseteq \overline{S}_{_N}(s_1)\dots\overline{S}_{_N}(s_n)\right\}=\sum_{g\in A_{_N}}
\sharp\left\{\underline{r}
\subseteq \overline{S}_{_N}(g)\right\}\sharp\left\{g\subseteq s_1^n\right\}\\ 
&+\sum_{p=2}^k\sum_{g_1,\dots,g_p\in A_{_N}}\sharp\left\{\underline{r}
\frown \overline{S}_{_N}(g_1)\dots\overline{S}_{_N}(g_p)\right\}\sharp
\left\{g_1\dots g_p\subseteq s_1^n\right\}
\end{array}
\end{equation}
where $\sharp\left\{\underline{r}
\frown \overline{S}_{_N}(g_1)\dots\overline{S}_{_N}(g_p)\right\}$ is
the number of occurrences of $\underline r$ in the string
$\overline{S}_{_N}(g_{1})\dots \overline{S}_{_N}(g_{p})$ which start
in $\overline{S}_{_N}(g_{1})$ and end in $\overline{S}_{_N}(g_{p})$.
We obtain
\begin{equation}
\hskip-30pt
\begin{array}{ll}
\mu(\underline{r})&=\lim_{n\to\infty}\frac{n}{|\overline{S}_{_N}(s_1^n)|}
\left(\sum_{g\in A_{_N}}
\sharp\left\{\underline{r}
\subseteq \overline{S}_{_N}(g)\right\}\frac{\sharp
\left\{g\subseteq s_1^n\right\}}{n}\right.\\ \nonumber
&+\left.\sum_{p=2}^k\sum_{g_1,\dots,g_p\in A_{_N}}\sharp\left\{\underline{r}
\frown \overline{S}_{_N}(g_1)\dots\overline{S}_{_N}(g_p)\right\}
\frac{\sharp
\left\{g_1\dots g_p\subseteq s_1^n\right\}}{n}\right)\\ \nonumber
&= \frac{1}{\overline{Z}_{_N}}
\left(\sum_{g\in A_{_N}}
\sharp\left\{\underline{r}
\subseteq \overline{S}_{_N}(g)\right\}\mu_{_N}(g)\right.\\ \nonumber
&+ \left.\sum_{p=2}^k\sum_{g_1,\dots,g_p\in A_{_N}}\sharp\left\{\underline{r}
\frown \overline{S}_{_N}(g_1)\dots\overline{S}_{_N}(g_p)\right\}
\mu_{_N}(g_1\dots g_p)\right)
\end{array}
\end{equation}

Let $\mathcal{P}$ be the projection operator that maps a measure
$\mu$ to its $1$-Markov approximation $\mathcal{P}\mu$ and define
$\pi_{N}^j=\mathcal S_{_{j+1}}\dots \mathcal S_{_N}
\mathcal{P}\mu_{_N}$. In particular we have $\pi_{_N}^0 = \overline
{\mathcal S}_{_N}\mathcal{P}\mu_{_N}$ and
$\pi_{_N}^{^{N}}=\mathcal{P}\mu_{_N}$. It holds

\begin{equation}
\label{inv:markov}
\pi_{_N}^{^N} = \overline{\mathcal G}_{_N}\pi_{_N}^0.
\end{equation}
In fact the measures $\pi_{_N}^{^N}$ and $\mu_{_N}$ coincide on the
pairs of symbols, then $\pi_{_N}^{^N}(\underline w)=0$ if
$\underline w \notin \overline{G}_{_N}(A^*)$, as follows from Th.
\ref{teo:vincoli}. Being $\overline{G}_{_N}(A^*)\subseteq
G_{_N}(A^*_{_{N-1}})$ we can apply Proposition \ref{propo:invS},
obtaining
\begin{equation}
\nonumber
\pi_{_N}^{^N} = \mathcal G_{_N}\pi_{_N}^{^{N-1}}.
\end{equation}
Now, also $\pi_{_N}^{^{N-1}}$ and $\mu_{_{N-1}}$ coincide on the
pairs of symbols (see Proposition \ref{prop:S}), then we can iterate
the procedure till to obtain Eq. (\ref{inv:markov}). Note that
\begin{equation}
\label{eq:zuguali}
\overline{Z}_{_N}^{\pi^0_{_N}}=\prod_{j=1}^N (1+\pi_{_N}^j(\alpha_j))=
\prod_{j=1}^N (1+\mu_j(\alpha_j)) = \overline{Z}_{_N}^{\mu},
\end{equation}
in fact $\pi_{_N}^j$ and $\mu_j$ coincide on the pairs of symbols on
$A_j$.
\noindent
Therefore
for any $k$ and any $\underline r$ of length $k$:
\begin{equation}
\label{Puteana2}
\hskip-40pt
\begin{array}{ll}
&|\pi_{_N}^0(\underline r)-\mu(\underline r)|\le\\ 
&\le \frac 1{\overline{Z}_{_N}} \sum_{p=3}^k \sum_{g_1,\dots g_p \in
A_{_N}} \Big(\mu_{_N}+
\pi_{_N}\Big)(g_{1}\dots g_{p})
\sharp\left\{\underline{r}
\frown \overline{S}_{_N}(g_1)
\dots\overline{S}_{_N}(g_p)\right\} \\
&\le 2\frac {k^2}{\overline{Z}_{_N}}
\end{array}
\end{equation}
which tends to 0 when $N\to +\infty$.
This implies that
$$\lim_{N\to +\infty} h_k(\pi_{_N}^{0}) = h_k(\mu).$$
In conclusion, for any $k$
$$
h(\mu)=
\frac {h(\mu_{_N})}{\overline{Z}_{_N}} \leq
\frac {h_1(\mu_{_N})}{\overline{Z}_{_N}} =
\frac {h(\pi_{_N}^{^N})}{\overline{Z}_{_N}} =
h(\pi_{_N}^0)\leq h_k(\pi_{_N}^0).
$$
We stress that the third step of the previous chain
follows from the definition $\pi_{_N}^{^N} = \mathcal{P} \mu_{_N}$
and that
the fourth step
follows from (\ref{inv:markov}) and (\ref{eq:zuguali}).

\noindent
Taking the limits $N\to +\infty$ and $k\to +\infty$:
$$h(\mu) =  \lim_{N\to +\infty} \frac {h_1(\mu_{_N})}{\overline{Z}_{_N}}.$$

\section{Some examples}
\label{sez:esempio}
We consider here a given sequence of pair substitutions
which is not obtained with the procedure
of the minimization of the length of the new strings, as prescribed
in the NSRPS method.

The initial alphabet is $A=\{0,1\}$. The first pair substitution is
$10\to 2$, the second $20\to 3$; in general the $N$-th  substitution
is $N0 \to N+1$. Notice that the infinite composition of these
substitutions corresponds to the coding procedure that substitute
maximal blocks of $k$ consecutive zeros, and the one that precedes
them, with the new symbol $k+1$.

If the initial measure gives positive probability to the pair $11$,
then the normalization can not diverge, namely
for an initial (typical) string of length $L$, after the transformations
there remain at most $\mu(11)L$ symbols.

\vskip5pt
Let us notice that only the first substitution involves
the symbol one, then it is easy to do the following computations:
\begin{eqnarray}
\nonumber
\mu_{_N} (1|1) &= \mu_{_1}(1|1) = \frac {\mu(11) - \mu(110)}{\mu(1)
- \mu(10)} =
\mu(1|11),\\
\nonumber
\mu_{_N} (1|11) &= \mu_{_1}(1|11) = \frac {\mu(111) -
\mu(1110)}{\mu(11) - \mu(110)} = \mu(1|111).
\end{eqnarray}
If for the initial measure $\mu(1|111) \neq \mu(1|11)$, then
$\mu_{_N}(1|11) \neq \mu_{_N}(1|1)$ for any $N$ and
$h_1(\mu_{_N})/\overline{Z}_{_N}$ can not converge to $h(\mu)$ (the
limiting process can not be a 1-Markov process).

\vskip5pt On the other hand we can consider as initial measure a
finite mean renewal process,  that is a stationary process for which
the distances between consecutive ones are i.i.d. random variables
with distribution $\left\{p_k\right\}_{k\geq 1}$ and
$E^0=\sum_{j=1}^{\infty}jp_j<\infty$. The entropy of such a process
is
$$
h(\mu)=\frac{-\sum_{k=1}^{\infty}p_k\log p_k}{E^0}
$$
An explicit computation of the marginals of $\mu_{_N}$ is
not difficult. It follows that
$$
Z_{_N}=Z^{\mu_{_{N-1}}}_{N0}=\frac{E^{^{N-1}}}{E^{^N}},\ \ \ \ \ \ \
\overline{Z}_{_N}=
\frac{E^{0}}{E^1}\frac{E^{1}}{E^2}\dots\frac{E^{^{N-1}}}{E^{^N}}=
\frac{E^{0}}{E^{^N}},
$$
where $E^{^N}=\sum_{j=1}^{\infty}jp^{^N}_j$ and
$$
p^{^N}_j=\left\{
\begin{array}{ll}
p_1+\dots +p_{N+1} & j=1\\
p_{N+j} & j>1\\
\end{array}
\right.
$$
Note that if we  consider the measures $\mu_{_N}$ as measures in the
alphabet $\mathbb N$, then $\mu_{_N}$ weakly converges to the
product measure with marginals $\left\{p_k\right\}_{k\geq 1}$. From
this (or by direct computation) we can derive
$$
\lim_{N\to \infty}\frac{h_1(\mu_{_N})}{\overline{Z}_{_N}}=
\lim_{N\to \infty}\frac{H_1(\mu_{_N})}{\overline{Z}_{_N}}=
h(\mu)
$$
Let us stress that in this case the process becomes independent,
then also $H_1(\mu_{_N})/\overline Z_{_N}$ converges to the entropy.
This fact is a consequence of the very particular choice of the
initial measure. If the distances between consecutive ones are not
distributed independently, but, for instance, with a two step Markov
process, then $h_1(\mu_{_N})/{\overline{Z}_{_N}}$ and
$H_1(\mu_{_N})/ \overline {Z}_{_N}$ do not converge to the entropy.

\section{Concluding remarks}
\label{sez:concluding}

The main result proved here says that under the action of the NSRPS
procedure any ergodic process becomes asymptotically Markov, i.e.
$h_1(\mu_{_N})/\overline{Z}_{_N}\to h$. A natural question is when
the process becomes even independent, i.e.
$H_1(\mu_{_N})/\overline{Z}_{_N}\to h$, as for the very specific
example discussed in section 5. In our opinion this is a non trivial
question, presumably depending on the behavior of the number of
forbidden sequences in the iterated measures.

The results of this paper imply also the fact that a NSRPS algorithm
can be used to estimate the entropy of an ergodic source starting
from a sequence of sufficiently large length, say $L$. This is done
iterating $N(L)$ pair substitutions with $N(L)$ diverging with $L$
sufficiently slow, and then computing the conditional entropy $h_1$
of the empirical measure of the resulting sequence. An interesting
question is how fast $N(L)$ can diverge with $L$.

Analogously it is possible to define an asymptotically optimal
compression algorithm based on NSRPS: iterating a suitable number of
times the pair substitution procedure we ends up with an
approximatively Markov sequence; this sequence can be compressed by
an algorithm which take into account only the pair correlations
(like for instance a suitable arithmetic coder). As before, if the
number of substitutions diverges with $L$ sufficiently slow then the
compression rate converges to $h$.

In practice, given a sequence of length $L$, it is not so obvious to
decide in an efficient way what is the optimal number of
substitution to make. This point is discussed a little bit in
\cite{G} and we do not enter in it.

\section{Technical results on measures transformations}
\label{appendice:mis}
\subsection{Proof of Theorem \ref{teo:mis} }
We do not give a formal proof of the theorem but just a sketch
of it (more details are in the analogous proof for
Proposition \ref{teo:ZW}, in the next subsection).
The fact that the limits are almost surely constants
can be deduced from the strong law of large numbers.
This fact implies the ergodicity of $\mathcal{G}\mu$ and $\mathcal{S}\nu$
(see theorem I.4.2 pag. 44 of \cite{S}). The compatibility conditions
for the families of marginals are easily checked.
Formula
(\ref{misSG}) is consequence of (\ref{mapSG}).

\subsection{Proof of Proposition \ref{teo:ZW}}

\noindent
In the case $x\neq y$, we have
$$
|G(w_1^n)|=n-\sharp\left\{xy\subseteq w_1^n\right\}
$$
so that
$$
\frac{n}{|G(w_1^n)|}=\frac{1}{1-
\frac{\sharp\left\{xy\subseteq w_1^n\right\}}{n}}
$$
and the result (\ref{def:Z})
follows from the strong law of large numbers.

\noindent
In the case $x=y$ we have that
$$
|G(w_1^n)|=n-\sum_{k=2}^n\sharp\left\{* x^k *
\subseteq w_1^n\right\}\left[\frac k2\right]
$$
where $\left[\ \right]$ is the integer part and
$\sharp\left\{* x^k *
\subseteq w_1^n\right\}$ is the number of blocks of exact length k
of consecutive $x$ contained in $w_1^n$ ($*$ represent
a possible occurrence of
a generic letter different from $x$).
It holds
$$
\sharp\left\{* x^k *
\subseteq w_1^n\right\}=\sharp\left\{x^k
\subseteq w_1^n\right\}-2\sharp\left\{x^{k+1}
\subseteq w_1^n\right\}+\sharp\left\{x^{k+2}
\subseteq w_1^n\right\}
$$
Now we have
$$
\frac{n}{|G(w_1^n)|}=\frac{1}{1-\sum_{k=2}^n(-1)^k\left(\frac{\sharp\left\{x^k
\subseteq w_1^n\right\}}{n}\right)}
$$
that converges to the right hand side of (\ref{def:Zxx}) for any
ergodic measure $\mu$ different from
%$\delta_{\underline{x}}$, the delta
the measure concentrated on the sequence of all $x$ (in this case
clearly $Z=2$).

Formula (\ref{def:W}) follows from
$$
S(z_1^n)=n+\sharp\left\{\alpha \subseteq z_1^n\right\}
$$
and the strong law of large numbers.

Formula (\ref{Z=W}) can be deduced from (\ref{mapSG}).

\subsection{$\mathcal S\nu$ in terms of $\nu$}
\label{sub:snu} We consider the substitution $\alpha \to xy$. We
have that
$$W = \lim_{n\to +\infty} \frac{|S(z_1^n)|}{n}
=  \lim_{n\to +\infty} \sum_{|\underline z|=n} \nu(\underline z)
 \frac{|S(\underline z)|}{n},
$$
it holds
\begin{equation}
\label{e1}
\begin{array}{ll}
\mathcal S\nu(\underline r) &:=
\lim_{n\to +\infty}
\frac {\sharp \{\underline r \subseteq S(z_1^n) \}}
{|S(z_1^n)|} =
\lim_{n\to +\infty} \frac {\sharp \{\underline r
\subseteq S(z_1^n) \}}{Wn}\\
&\phantom{:}=\lim_{n\to +\infty} \frac 1{Wn} \sum_{|\underline z|=n} \nu(\underline z)
\sharp \{\underline r
\subseteq S(z_1^n) \}.
\end{array}
\end{equation}
Suppose now that  $|\underline r|=k$ and
consider, for $n\geq k$,
\begin{equation}
\label{e2}
\hskip-30pt
\begin{array}{ll}
D_n &= 
\sum_{|\underline z|=n} \nu(\underline z)\sharp \{\underline r
\subseteq S(\underline z) \} -
\sum_{|\underline z|=n-1}  \nu(\underline z)\sharp \{\underline r
\subseteq S(\underline z) \} \\
&= \sum_{|\underline z|=n}  \nu(\underline z) \id\left( \underline{r}
= S(\underline z)_1^k \right) +
 \sum_{|\underline z|=n-1}  \nu(\alpha \underline z)
 \id\left(
\underline{r} = y S(\underline z)_1^{k-1} \right).
\end{array}
\end{equation}
We can rewrite these terms as
\begin{equation}
\hskip-40pt
\begin{array}{ll}
&\sum_{|\underline z|=n} \nu(\underline z) \id\left(
r = S(\underline z)_1^k \right) =
\sum_{\underline s: S(\underline s) =
\underline r} \left( \nu(\underline s) +
\nu(s_1^{|\underline s|-1}\alpha) \id \left(r_k = x \right) \right)\\
\label{e3}
&\sum_{|\underline z|=n-1}  \nu(\alpha \underline z)
\id\left(r = y S(\underline z)_1^{k-1} \right) = \\
&\phantom{aaaaaaaaaaa}
= \sum_{\underline s: S(\underline s) = \underline r}
\left( \nu(\alpha s_2^{|\underline s|})
+  \nu(\alpha s_2^{|\underline s|-1}\alpha) \id( r_k = x)
\right) \id(r_1 = y).
\end{array}
\end{equation}
Hence $D_n$ is constant for $n\geq k$ and
\begin{eqnarray}
\label{e4}
\lim_{n\to +\infty}
\frac 1W \sum_{|\underline z|=n} \nu(\underline z)
\sharp \{\underline r
\subseteq S(z_1^n) \} = \frac 1W D_k
\end{eqnarray}
Collecting (\ref{e1})--(\ref{e4}) we obtain
the expression for $\mathcal S\nu$:
\begin{equation}
\label{espr:S}
\hskip-30pt
\begin{array}{ll}
\mathcal S\nu(\underline r) = \frac 1W &
\sum_{\underline s:\,S(\underline s)= \underline r}
\left( \nu(\underline s) +
\nu(s_1^{|\underline s|-1}\alpha) \id \left(r_k = x \right) +\right.\\
&\left. \nu(\alpha s_2^{|\underline s|}) \id(r_1 = y) +
 \nu(\alpha s_2^{|\underline s|-1}\alpha )\id(r_1 = y) \id( r_k = x)
\right)
\end{array}
\end{equation}

\subsection{$\mathcal G\mu$ in terms of $\mu$}
\label{sub:gmu}
The map $S$ inverts $G$, then in order to find the expression
of $\mathcal G\mu$ we can invert the expression of $\mathcal S \mathcal
G\mu = \mu$. Let $\nu$ be $\mathcal G\mu$. The sum on $\underline s$
in Eq. (\ref{espr:S}) reduces  to $\underline s = G(\underline r)$,
namely $\nu(\underline s)=0$ if $\underline s \notin G(A^*)$.
This reduction make explicitly invertible Eq. (\ref{espr:S}),
but we have to distinguish the two cases $x\neq y$ and $x=y$.

\vskip5pt
\noindent
{\bf Case $x\neq y$}.

Let $\underline r\in A^*$ and
let $z,\,w\in A$ be such that $z\neq x$ and $w\neq y$. From (\ref{espr:S})
we obtain:
\begin{equation}
\label{espr:Sparicolare}
\hskip-40pt
\begin{array}{ll}
&W\mu(w\underline r z ) = \nu(G(w\underline rz)) \\
&W\mu(w\underline r x ) =  \nu(G(w\underline r)x)
+  \nu(G(w\underline r)\alpha ) \\
&W\mu(y\underline r z ) = \nu(yG(\underline rz))
+ \nu(\alpha  G(\underline rz))\\
&W\mu(y\underline r x ) = \nu(yG(\underline r)x)
+ \nu(\alpha  G(\underline r)x)
+ \nu(yG(\underline r)\alpha )
+ \nu(\alpha  G(\underline r )\alpha)
\end{array}
\end{equation}
Let now $\underline s = G(\underline r)$ with
$|\underline s|=n$  and $|\underline r|=k$.
The expression of $\nu(\underline s)= \mathcal G\mu (\underline s)$
can be calculated from
the previous equations, obtaining
%\vskip5pt

\begin{equation}
\label{inv1}
\hskip-40pt
\begin{array}{l l l}
s_1\neq y, & s_n \neq x: & \nu(\underline s) = W\mu(\underline r)\\
s_1 = y, &  s_n \neq x: & \nu(\underline s) = W(\mu(\underline r)
-\mu(xyr_2^k))\\
s_1 \neq y, & s_n = x: & \nu(\underline s) =W(\mu(\underline r)-
\mu(r_1^{k-1}xy)\\
s_1 = y, & s_n = x: &  \nu(\underline s) =W(\mu(\underline r)
+\mu(xyr_2^{k-1}xy) -
\mu(xyr_2^k) - \mu(r_1^{k-1}xy))
\end{array}
\end{equation}

%\begin{tabular}{l | l | l }
%  if $s_1$ is & if $s_k$ is &  $\mathcal G\mu (\underline s) =
%\nu(\underline s)$ is\\
%\hline
%$\neq y$ & $\neq x$ & $W\mu(\underline r)$ \\
%$= y$ & $\neq x$ & $W(\mu(\underline r)
%-\mu(xyr_2^k))$ \\
%$\neq y$ & $=x$ & $W(\mu(\underline r)-
%\mu(r_1^{k-1}xy)
%$ \\
%$=y$ & $=x$ & $W(\mu(\underline r)
%+\mu(xyr_2^{k-1}) -
%\mu(xyr_2^k) - \mu(r_1^{k-1}xy))
%$
%\end{tabular}

\vskip5pt
\noindent
Now we can calculate $Z=W$ (see Eq. (\ref{Z=W})) in terms of $\mu$:
$$Z=1 + \nu(\alpha) = 1 + Z\mu(xy) = \frac 1{1-\mu(xy)}.$$
We remark that eq.s (\ref{inv1}) can be synthesized in
\begin{equation}
\label{inv1bis} \mathcal G\mu(\underline s) = Z \sum_{a,b\in A:\,
a\underline{s}b \in G(A^*)} \mu(a\underline{s}b).
\end{equation}

\vskip5pt
\noindent
{\bf Case $x=y$}.

Proceeding as above we obtain again the explicit expressions for
$\nu(\underline s)$ but they are more complicated.
As before let $\underline s\in G(A^*)$, $|\underline s|=n>0$,
$G(\underline r) = \underline s$, $|\underline r|=k$.
Let $s_1,s_n \neq x$.
Denoting with $\underline a^p$ the string of
$p$ times the symbol $a$,  the strings in $G(A^*)$ are
of the type
$$\underline \alpha^p
\underline x^{\pi}
\underline s\,\underline \alpha^q\underline x^{\sigma}\ \text{ and }
\ \underline \alpha^p\underline x^{\pi},\
\text{ with }p,q\geq 0\text{ and }\pi,\sigma = 0,\,1.
$$
The expression of $\mathcal G\mu=\nu$ in terms of $\mu$ is given by:
\begin{equation}
\label{inv2}
\hskip-60pt
\begin{array}{l l l}
\nu(\underline s\,\underline \alpha^q) &= Z \mu(\underline r\,\underline
x^{2q})&\text{for }q\geq 0\\
\nu(\underline s\,\underline \alpha^q x) &= Z (\mu(\underline r\,\underline
x^{2q+1}) - \mu(\underline r\,\underline
x^{2q+2})))&\text{for }q\geq 0\\
\nu(\underline \alpha^p) &= Z\sum_{j=0}^{+\infty}(-1)^j
\mu(\underline x^{2p+j}))&\text{for }p> 1\\
\nu(\underline \alpha^px) &= Z(
\mu(\underline x^{2p+1}) -
2\sum_{j=2}^{+\infty}(-1)^j
\mu(\underline x^{2p+j}))&\text{for }p> 1\\
\nu(\underline \alpha^p\underline x^{\pi}\underline s
\,\underline \alpha^q) &=
Z\sum_{j=0}^{+\infty}(-1)^j
\mu(\underline x^{2p+\pi+j}\underline r\,\underline x^{2q})
&\text{for }p+\pi\geq 1,\,q\geq 0\\
\nu(\underline \alpha^p\underline x^{\pi}\underline s\,
\underline \alpha^qx) &=
Z\sum_{j=0}^{+\infty}(-1)^j\cdot &
\\&(\mu(\underline x^{2p+\pi+j}\underline r\,\underline x^{2q+1})
-\mu(\underline x^{2p+\pi+j}\underline r\,\underline x^{2q+2}
))&\text{for }p+\pi\geq 1,\,q\geq 0
\end{array}
\end{equation}
Now we can calculate $Z$ in terms of $\mu$:
$$Z = 1 + \nu(\alpha) = 1 +
Z \sum_{j=0}^{+\infty}
(-1)^j \mu(\underline x^{2+j}) =
\frac 1{1-\sum_{j=2}^{+\infty} (-1)^j \mu(\underline x^j)}
$$

\subsection{Proof of Proposition \ref{prop:S}}
This proposition is a consequence of Eq. (\ref{espr:S}) in
subsection \ref{sub:snu},
namely $|\underline s|\leq \underline r$
if $S(\underline s)=\underline r$.

\subsection{Proof of Proposition \ref{propo:invS}}
This proposition is a consequence of the fact that the explicit
expression (\ref{espr:S}) of $\mu=\mathcal{S}\nu$ in terms of $\nu$
can be inverted (in an unique way) if $\nu$ respects the pair
constraints given by $G$, as follows from eq.s
(\ref{espr:Sparicolare})--(\ref{inv2}) in subsection \ref{sub:gmu}.
The expression of $\nu$ in terms of $\mu$ is exactly $\mathcal
G\mu$, then $\nu=\mathcal G\mu = \mathcal G \mathcal S \nu$.

\section{Technical results on entropies transformations}
\label{appendice:entropie}

\subsection{Proof of theorem \ref{teo:h}}

The result follows from the fact that $G$ is a faithful code and $S$
is a faithful code when restricted to the support of
$\mathcal{G}\mu$. We call $C:=\left\{C_n\right\}_{n\in \mathbb N}$ a
sequence of universal codes in the alphabet $A$ and
$C':=\left\{C'_n\right\}_{n\in \mathbb N}$ a sequence of universal
codes in the alphabet $A'$ (see theorems II.1.1 and II.1.2 page 122
of \cite{S}).

We have that $C'\circ G$ is a sequence of
faithful codes in $A$. From this we deduce that on a set
of $\mu$ measure one
$$
h(\mathcal{G}\mu)=\lim_{n\to \infty}\frac{C'_{|G(w_1^n)|}(G(w_1^n))}
{|G(w_1^n)|}=\lim_{n\to \infty}\frac{n}{|G(w_1^n)|}
\frac{C'_{|G(w_1^n)|}\circ G(w_1^n)}{n}\geq Zh(\mu)
$$

Likewise we have that $C\circ S$ is a sequence of
faithful codes in $A'$. From this we deduce that on a set
of $\mu$ measure one
$$
h(\mu)=\lim_{n\to \infty}\frac{C_n(w_1^n)}{n}=
\lim_{n\to \infty}\frac{|G(w_1^n)|}{n}
\frac{C_{n}\circ S(G(w_1^n))}{|G(w_1^n)|}\geq \frac{h(\mathcal{G}\mu)}{Z}
$$

\subsection{Proof of theorems \ref{teo:h1} and \ref{teo:hk} }
We
%do not use
%the explicit expressions of $\mathcal G\mu$ given in
%appendix \ref{appendice:mis}, which are not useful for this purpose.
%We
proceed splitting the action of $G$  (and then of $\mathcal G$)
%, on a given string,
in three parts, introducing two new character $b_1$, $b_2\notin A$.

\noindent
Given a string, we operate as follows:

\begin{enumerate}[Step 1 ]
\item We substitute, starting form the left, any occurrence
of $xy$ with $xb_1$. This operation define a map
$R:A^*\to A_{R}^*$, where $A_{R} = A\cup \{b_1\}$.
We call $\mathcal R$
the corresponding map for the measures, defined in the same spirit
of Th. \ref{teo:mis}.
\item
We substitute any occurrence of $xb_1$ with $b_2b_1$.
This operation define a map
$L:A_{R}^*:\to A_{L}^*$,
 where $A_{L} = A_{R}\cup \{b_2\}$.
We call $\mathcal L$
the corresponding map for the measures.
\item
We substitute any occurrence of $b_2b_1$ with $\alpha$.
This operation, in general, define a map
$C:A_{L}^*:\to A_{C}^*$,
 where $A_{C} = A_{L}\cup \{\alpha\}$.
We call $\mathcal C$ the corresponding map for the measures.
\end{enumerate}

\vskip8pt
\noindent
From these definitions:
$$C(L(R(\underline w)))= G(\underline w),\ \ \text{ and then }\
\mathcal C \mathcal L \mathcal R \mu = \mathcal G \mu.$$
With this splitting we separate the effects of the shortening of the
strings (step 3) from the effect of the partial substitutions of characters
(steps 1, 2).

\begin{lemma}
\noindent
\label{lemma1}
\begin{equation}
\label{passo1}
h_1(\mathcal R\mu) \leq h_1(\mu)
\end{equation}
\end{lemma}
(the proof is in subsection \ref{sub:lemma8.1}).

The same assertion holds for $\mathcal L\mathcal R \mu$. Namely
we can define $L$ also considering the substitutions starting from the right,
namely $x\neq b_1$. In this way
$L(\underline w) = ( R'(\underline w^r))^r$,
where $\underline w^r=(w_1\dots w_k)^r=w_k\dots w_1$ and
$R'$ is the substitution, from the left, of $b_1x$ with $b_1b_2$.
The map $R'$ acts in the same way of $R$,
then lemma \ref{lemma1} holds
for the corresponding map for the measures $ {\mathcal R}'$, and
then also for $\mathcal L$. In this way we prove that
$$h_1(\mathcal L\mathcal R\mu) \leq h_1(\mu).$$

The third step preserves $h_1$ up to the normalization, as stated in
the following lemma (proved in subsection \ref{sub:lemma8.2})
\begin{lemma}
\label{lemma2}
If $\rho \in \mathcal E(A_{L})$ verifies
\begin{equation}
\label{vincolob1b2}
\rho(b_2w)=\rho(zb_1)=0\ \ \text{ for } \ \ w\neq b_1,\, z\neq b_2,
\end{equation}
then
\begin{equation}
\label{passo3}
h_1(\mathcal C\rho) = W h_1(\rho),
\end{equation}
where
\begin{equation}
W = \frac 1{1-\rho(b_2b_1)} = 1 + \mathcal C\rho(\alpha).
\end{equation}
\end{lemma}

\vskip5pt \noindent We achieve the proof of theorem \ref{th:teoh1}
observing that the measure $\rho =  \mathcal L \mathcal R \mu$
verifies the constraints (\ref{vincolob1b2}), then $h_1(\mathcal
G\mu )\leq W h_1(\mu)$, where $W=Z$ because $W=1+\mathcal C\rho
(\alpha) = 1 + \mathcal G\mu(\alpha)= W_{\alpha}^{\mathcal
G\mu}=Z_{xy}^{\mu}$ (see Eq. (\ref{Z=W})).

\vskip10pt
We conclude this section remarking that
Lemma \ref{lemma1} holds also for $h_k$, and that for $h_k$ it holds the
following analogous of
Lemma \ref{lemma2}, proved in subsection \ref{sub:lemma8.3}
\begin{lemma}
\label{lemma3} Under the hypotheses of lemma \ref{lemma2}
$$h_k(\mathcal C \rho) \leq W h_k(\rho)$$
\end{lemma}
From these facts it follows Theorem \ref{teo:hk}.

\subsection{Proof of Lemma \ref{lemma1}}
\label{sub:lemma8.1}
Let $\xi=\mathcal R\mu$.
The measure $\mu$ can be expressed in terms of $\xi$ as follows:
$$\mu(\underline w) = \sum_{\underline z:\,R(\underline z) =
\underline w} \xi(\underline z).$$
We use this formula
to express the probabilities of the symbols and of the pairs of symbols.

\noindent
{\bf Case $x\neq y$}. Let $p$ be in $A$:
$$
\begin{array}{l ll}
\mu(y) = \xi(y) + \xi(b_1), & \mu(p) = \xi(p)\ &\text{ for } p\neq y, \\
\mu(yp) = \xi(yp) + \xi(b_1p), \ \ \ &
\mu(pq) = \xi(pq)\ &\text{ for }p\neq x,\,p\neq y,\\
\mu(xy) = \xi(xb_1), & \mu(xp) = \xi(xp)\ &\text{ for }p\neq y. \\
\end{array}$$
By direct calculation:
\begin{equation}
\label{eq:diffh1}
\begin{array}{ll}
h_1(\mu) - h_1(\xi) =
&-\sum_{p\in A} (\xi(yp)+\xi(b_1p))\log
\frac {\xi(yp)+\xi(b_1p)}{\xi(y)+\xi(b_1)}\\
&+\sum_{p\in A} \left(
\xi(yp) \log \frac {\xi(yp)}{\xi(y)}+
\xi(b_1p) \log \frac {\xi(b_1p)}{\xi(b_1)}\right)
\end{array}
\end{equation}
We prove the lemma showing that:
$$
\begin{array}{l}
\left(
\xi(yp) \log \frac {\xi(yp)}{\xi(y)}+
\xi(b_1p) \log \frac {\xi(b_1p)}{\xi(b_1)}\right)\geq
(\xi(yp)+\xi(b_1p))\log
\frac {\xi(yp)+\xi(b_1p)}{\xi(y)+\xi(b_1)}.
\end{array}
$$
Dividing for $\xi(yp)+\xi(b_1p)$ and setting
$\beta = \frac{\xi(y)}{\xi(y) + \xi(b_1)}$,
$\gamma = \frac{\xi(yp)}{\xi(yp)+\xi(b_1p)}$,
this inequality can be rewritten as:
$$\gamma \log \frac {\beta}{\gamma} +
(1-\gamma) \log \frac {1-\beta}{1-\gamma}\leq 0, $$
which is always verified.

\vskip5pt
\noindent
{\bf Case $x=y$}. Let $p\in A$, $p\neq x$:
$$
\begin{array}{l l}
\mu(x) = \xi(x) + \xi(b_1), & \mu(p) = \xi(p),\\
\mu(xx) = \xi(xb_1) + \xi(b_1x),\ \  &
\mu(xp) = \xi(xp) + \xi(b_1p),\\
\mu(pq) = \xi(pq)\ \ \text{ for }q\in A,
& \mu(px) = \xi(px).\\
\end{array}$$
The difference between the  1--conditional entropies is
\begin{equation}
\hskip-30pt
\begin{array}{ll}
h_1(\mu) - h_1(\xi) =
&-\sum_{p\in A,\,p\neq x} (\xi(xp)+\xi(b_1p))\log
\frac {\xi(xp)+\xi(b_1p)}{\xi(x)+\xi(b_1)}\\
&- \left( \xi(xb_1) + \xi(b_1 x)\right) \log \frac
{ \xi(xb_1) + \xi(b_1 x)}{\xi(x)+\xi(b_1)}\\
&+\sum_{p\in A,\,p\neq x} \left(
\xi(xp) \log \frac {\xi(xp)}{\xi(x)}+
\xi(b_1p) \log \frac {\xi(b_1p)}{\xi(b_1)}\right)
\\
&+\xi(xb_1) \log \frac {\xi(xb_1)}{\xi(x)}
+\xi(b_1x) \log \frac {\xi(b_1x)}{\xi(b_1)}
\end{array}
\end{equation}
We prove that this difference is positive
with the same argument used for the case $x\neq y$.

\vskip5pt
\noindent
Finally, we remark that in the same way we can prove
that $h_k(\xi)\leq h_k(\mu)$.

\subsection{Proof of Lemma  \ref{lemma2}}
\label{sub:lemma8.2}
Let $\nu=\mathcal C\rho$ and $W=1+\nu(\alpha)$.
It is easy to write $\rho$ in terms of $\nu$. Let $p,q\neq b_1,b_2$.
The probabilities of the symbols and of the
pairs of symbols are given by

$$
\begin{array}{l l}
W\rho(b_1) = W\rho(b_2) = \nu(\alpha) \ \ \ & W\rho(p) = \nu(p)  \\
W\rho(pb_1) = W\rho(b_2q) =0 &
W\rho(pq) = \nu(pq)\\
W\rho(pb_2) = \nu(p\alpha) &
W\rho(b_1q) = \nu(\alpha q)\\
W\rho(b_2b_1) = \nu(\alpha) &
W\rho(b_1b_2) = \nu(\alpha\alpha)
\end{array}
$$
By explicit calculation:
$$
\begin{array}{l l}
H_1(\rho) &= -\sum_{p\in A_{C}\backslash \alpha}
\frac {\nu(p)}W \log \frac {\nu(p)}W - 2 \frac {\nu(\alpha)}W
\log \frac {\nu(\alpha)}W
= \frac {H_1(\nu)}W + \frac{\log W}W
-\frac {\nu(\alpha)}W \log \frac {\nu(\alpha)}W,\\
H_2(\rho) &= -\sum_{p,q\in A_{C}} \frac {\nu(pq)}W \log \frac {\nu(pq)}W -
\frac {\nu(\alpha)}W \log \frac{\nu(\alpha)}W
= \frac {H_2(\nu)}W +\frac {\log W}W -
\frac {\nu(\alpha)}W \log \frac{\nu(\alpha)}W.
\end{array}
$$
Then
$$h_1(\rho)= H_2(\rho) - H_1(\rho) = \frac {h_1(\nu)}W.$$

\subsection{Proof of Lemma \ref{lemma3}}
\label{sub:lemma8.3}

We need some  definitions.
Let $\underline w=w_1^l$.
We can identify $\underline{w}$ with the cylindrical
subset  $K_{\underline w}\subseteq A^{\mathbb Z}$ defined as follows
$$
K_{\underline w}=\left\{\underline{x}\in A^{\mathbb Z}:
x_{-l}=w_1,x_{-l+1}=w_2,\dots ,x_{-1}=w_{l}\right\}
$$
Let $P\subseteq A^*$ be a finite set. We say that $P$ is a partition if
$$\{ K_{\underline w} \}_{\underline w \in P}
\ \ \text{ is a partition of }A^{\mathbb Z},
\text{ i.e. }
\left\{\begin{array}{l l}
(1)\ \ \ &K_{\underline w} \cap  K_{\underline z}=
\emptyset\ \text{ if }\underline w \neq \underline z, \\
(2)\ \ \ &\bigcup_{\underline w\in P} K_{\underline w} = A^{\mathbb Z}.
\end{array}
\right.
$$
Condition $(1)$ says that any string of $P$ is not a suffix for
others strings of $P$. If only condition $(1)$ is verified, we say
that $P$ is a semi-partition. It is easy to show that any
semi-partition can be completed to obtain a partition. Moreover, if
the minimum of the length of the strings in $P$ is $l$, we can
complete $P$ using strings of length greater or equal to $l$.

If $P$ is a partition, we can define the $P$-conditional entropy
as
$$h_P(\mu)=-\sum_{\underline w\in P,\,a\in A}
\mu(\underline wa) \log \frac
{\mu(\underline wa)}{\mu(\underline w)}.$$
If $P$ and $Q$ are two partition we say that
$P$ is more fine of $Q$ if any string of $P$ ends with a
string of $Q$.
If $P$ is more fine then $Q$:
\begin{equation}
\label{piufine}
h_P(\mu) \leq h_Q(\mu).\end{equation}
(The proof is at the end of this section).

Note that
$$P=\{\underline s \in A_{L}^*|\, |C(\underline s)| = k\},$$
is a semi-partition, and that, from direct calculation
$$h_k(\nu) = W h_P(\rho).$$
Where we remember that $\nu={\mathcal C}\rho$. In particular we have
used that, if $\underline s\in A_{L}^*$, $\rho(\underline s b_2) =
\rho(\underline s b_2b_1)$ and if the last symbol of $\underline s$
differs from $b_2$ then $\rho(\underline sb_1)=0$.

Finally let $\overline P$ be a completion of $P$.
$$h_k(\nu) = W h_P(\rho) \leq W h_{\overline P}(\rho).$$
The length of the strings in $P$ is greater or equal to $k$ and we
construct $\overline{P}$ so that the same holds for $\overline P$.
Therefore, $A_{L}^k$ is a partition less fine of $\overline P$.
Invoking Eq. (\ref{piufine}) we conclude that
$$h_k(\nu) \leq W h_{A_{L}^k}(\rho)= Wh_k(\rho).$$

\vskip5pt
\noindent
{\bf Proof of Eq. (\ref{piufine})}.

Let $\underline w\in Q$ and $X_{\underline w}\subseteq P$ be
the subset of the strings which end with $\underline w$.
From this definition:
$$P=\bigcup_{\underline w\in Q}X_{\underline w},\ \ \
\mu(\underline w) = \sum_{\underline r\in X_{\underline w}}
\mu(\underline r).$$
The function $\Phi(x) = x\log x$ is convex, then if
$\lambda_i\geq 0 $ and $\sum \lambda_i = 1$,
$\Phi(\sum \lambda_i x_i)\leq \sum \lambda_i x_i \log x_i$.
Now
$$
\begin{array}{l}
-h_Q(\mu) =
\sum_{\underline w\in Q} \mu(\underline w) \sum_{a\in A}
\mu(a|\underline w) \log \mu(a|\underline w),
\end{array}
$$
and
$$
\begin{array}{l}
\mu(a|\underline w) =
\frac {\mu(\underline w a)}{\mu(\underline w)}=
\sum_{\underline r \in X_{\underline w}}
\frac {\mu(\underline r a)}{\mu(\underline w)}
= \sum_{\underline r \in X_{\underline w}}
\frac {\mu(\underline r a)}{\mu(\underline r)}
\frac {\mu(\underline r)}{\mu(\underline w)}.
\end{array}
$$
Indicating with
$x_{\underline r}^a={\mu(\underline r a)}/{\mu(\underline r)}$, and with
$\lambda_r =  {\mu(\underline r)}/{\mu(\underline w)}$
and noting that 
$\sum_{\underline{r}\in X_{\underline w}} \lambda_{\underline r} = 1$,
we obtain:
\begin{equation}
\begin{array}{ll}
-h_Q(\mu) &= \sum_{\underline w\in Q} \sum_{a\in A}
\mu(\underline w) \Phi\left(\sum_{\underline r\in X_{\underline w}}
\lambda_{\underline r}x_{\underline r}^a \right) \\
&\leq \sum_{\underline w\in Q}\sum_{\underline r\in X_{\underline w}}
\sum_{a\in A}
\mu(\underline w) \frac {\mu(\underline r)}{\mu(\underline w)}
\mu(a|\,\underline r) \log \mu(a|\,\underline r)
\\
&= \sum_{\underline r\in P}\sum_{a\in A} \mu(\underline r)
\mu(a|\,\underline r) \log \mu(a|\,\underline r) = -h_P(\mu)
\end{array}
\end{equation}

\section*{Acknowledgements}
We thank V. Loreto and G. Parisi for very useful discussions.
D.B. acknowledges the support of cofin MIUR. E.C. acknowledges the
support of GNFM-INdAM and cofin MIUR.
D.G. acknowledges the support of GNFM-INdAM and cofin prin prot. $2004028108\_005$.

\end{document}